\documentclass[12pt]{article}
\usepackage{a4}
\usepackage{amsbsy}

\newlength{\abstwidth}
\newcommand{\RPP}[3]{{\it Rept.\ Prog.\ Phys.\ }{\bf #1} (#3) #2}

\newcommand{\PLB}[3]{{\it Phys.\ Lett.\ }{\bf B#1} (#3) #2}
\newcommand{\PRC}[3]{{\it Phys.\ Rep.\ }{\bf C#1} (#3) #2}
\newcommand{\NPB}[3]{{\it Nucl.\ Phys.\ }{\bf B#1} (#3) #2}
\newcommand{\ZPC}[3]{{\it Z.\ Phys.\ }{\bf C#1} (#3) #2}
\newcommand{\CPC}[3]{{\it Comput.\ Phys.\ Commun.\ }{\bf #1} (#3) #2}
\newcommand{\hep}[1]{[{\rm hep-ph/#1}]}

\newcommand{\pT}{p_{{\mathrm T}}}
\newcommand{\kT}{k_{{\mathrm T}}}

\renewcommand{\d}{{\mathrm d}}
\newcommand{\e}{{\mathrm e}}
\newcommand{\f}{{\mathrm f}}
\newcommand{\g}{{\mathrm g}}
\newcommand{\p}{{\mathrm p}}
\newcommand{\q}{{\mathrm q}}
\newcommand{\s}{{\mathrm s}}
\renewcommand{\u}{{\mathrm u}}
\newcommand{\qbar}{\bar{\mathrm q}}
\newcommand{\sbar}{\bar{\mathrm s}}
\newcommand{\ubar}{\bar{\mathrm u}}
\newcommand{\E}{{\mathrm E}}
\newcommand{\M}{{\mathrm M}}

\newcommand{\bJ}{{\boldsymbol{J}}}
\newcommand{\bP}{{\boldsymbol{P}}}
\newcommand{\beq}{{\boldsymbol{=}}}
\newcommand{\bzero}{{\boldsymbol{0}}}
\newcommand{\bone}{{\boldsymbol{1}}}
\newcommand{\btwo}{{\boldsymbol{2}}}
\newcommand{\bthree}{{\boldsymbol{3}}}
\newcommand{\bminus}{{\boldsymbol{-}}}
\newcommand{\bplus}{{\boldsymbol{+}}}

\begin{document}

\pagestyle{empty}

\begin{flushright}
CERN--TH/99--28\\ 
hep-ph/9902243
\end{flushright}

\vspace{\fill}

\begin{center}
{\Large\bf 
Central production of mesons:\\[1ex]
Exotic states versus Pomeron structure}\\[1.8ex]
{\bf Frank E.\ Close${}^{a}$} 
\\[1.4ex]
and \\ [1.5ex]
{\bf Gerhard A.\ Schuler${}^{b}$} \\[1.2mm] 
Theoretical Physics Division, CERN, CH-1211 Geneva 23 \\
\end{center}

\vspace{\fill}
\begin{center}
{\bf Abstract}\\[2ex]
\begin{minipage}{\abstwidth}
We demonstrate that the azimuthal dependence of 
central meson production in hadronic collisions, when suitably binned,  
provides unambiguous tests of whether the Pomeron couples like a conserved 
vector-current to protons.  
We discuss the possibility of discriminating between $\q\qbar$ and 
glueball production in such processes. 
Our predictions apply also to meson production in tagged two-photon 
events at electron--positron colliders and to vector-meson
production in $\e\p$ collisions at HERA.
\end{minipage}
\end{center}

\vspace{\fill}
\noindent
CERN--TH/99--28\\
February 1999

\vspace{\fill}
\noindent
\rule{60mm}{0.4mm}

\vspace{0.1mm} 
\noindent
${}^a$ On leave from Rutherford Appleton Laboratory, 
Chilton, Didcot, Oxfordshire, OX11 0QX, UK;
supported in part by 
the EEC-TMR Programme, Contract NCT98-0169.
\\[1.5ex]
${}^b$ Heisenberg Fellow; 
supported in part by the EU Fourth Framework Programme
``Training and Mobility of Researchers'', Network 
``Quantum Chromodynamics and the Deep Structure of Elementary Particles'', 
contract FMRX-CT98-0194 (DG 12-MIHT). 
\clearpage
\pagestyle{plain}
\setcounter{page}{1} 

\section{Introduction}
The production of mesons in the central region 
of proton--proton collisions ($\p\p \to \p\p M$) via a gluonic Pomeron
has traditionally been regarded as a potential source of glueballs 
\cite{Robson}. 
However, well-established quark--antiquark ($\q\qbar$) mesons 
are also known to be produced and this has led to searches for a selection
mechanism that could help to distinguish among such states. 
As a result it has been discovered \cite{CK}
that the pattern of resonances
produced in the central region of double tagged $\p\p\to \p\p M$
depends on the vector {\em difference} $\vec{k}_{\perp} = 
\vec{q}_{1\perp} - \vec{q}_{2\perp}$ of the transverse momentum recoils 
$\vec{q}_{i\perp}$ of the final protons (even at fixed four-momentum
transfers $t_i = - q_i^2$). When this quantity 
$\kT = |\vec{k}_{\perp}|$ is small 
($\leq {\cal O}(\Lambda_{\mathrm QCD})$) all well-established $\q\qbar$
states were observed to be suppressed \cite{WA}
while the surviving resonances 
included enigmatic states such as $\f_0(1500)$, $\f_J(1710)$ 
and $\f_2(1910)$ that have variously been suggested to be glueballs 
or to reside on the (gluonic) Pomeron trajectory. At large $\kT$,
by contrast, $\q\qbar$ states are prominent.

However, these $\kT$ dependences for at least $0^-$ and $1^+$ 
production have been shown to arise if the Pomeron (or perhaps
a hard gluonic component that produces $M$ by $\g\g$ fusion) 
transforms as a conserved vector current \cite{Close}. 
In order to help
determine the extent to which the double tagged reaction 
$\p\p \to \p\p M$ depends on a vector production or the dynamical structure
of the meson $M$ (of spin $J$ and parity $P$), 
we develop the earlier analysis to all $J \leq 3$. 

While the $\kT$ phenomenon has turned out to be a sharp experimental 
signature, we shall propose here that the azimuthal $\phi$ dependence 
(between the two proton scattering-planes in the $\p\p$ c.m.s.)
provides a rather direct probe of dynamics. In particular, observation
of non-trivial $\phi$ dependences  requires the presence of non-zero helicity transfer 
by the diffractive agent (Pomeron, gluon, $\ldots$) \cite{Diehl}
and so the Pomeron cannot simply transform as having vacuum quantum numbers: 
a spin greater than zero is needed. 
We analyse here the simplest case, where the process is driven by the
fusion of two 
spin-$1$ currents. Imposing current conservation it immediately applies
to $\e^+\e^- \to \e^+\e^- M$ and, empirically, already exhibits features
seen in $\p\p \to \p\p M$. 
We find that current non-conserving and/or scalar contributions 
are needed to accommodate the data. 

At extreme energies where non-diffractive contributions are negligible,
we show the following properties for meson production in the central region. 
\hfill\\
1. The $\phi$ dependence of $0^-$ production provides a clear test
for the presence of a significant vector component of the production Pomeron, 
independent of the $t$ dependence.
Preliminary data on $\eta$ and $\eta'$ production confirm this. 
\hfill\\
2. The production of $1^+$ mesons reinforces this: The conserved
vector-current (CVC) hypothesis implies (i)
the cross section will tend to zero as $\kT \to 0$, and (ii)
$1^+$ mesons are produced dominantly in the helicity-one state. 
Both features are prominent in the data. 
\hfill\\
3. The $0^+$ cross section survives at small $\kT$ for the CVC hypotheses. 
Moreover, for $q_{i\perp} \ll M$, we must either observe a 
$\cos^2\phi$ distribution or a small (relative to $0^-$) cross section.  
However, at least one $0^+$ state (the $f_0(2000)$) appears to
be suppressed at small $\kT$. Unlike $0^-$ or $1^+$ production, the
production of $0^+$ will be particularly sensitive to a scalar and/or 
non-conserved vector component to the Pomeron. 
In particular the vanishing of $f_0(1500)$ as $\phi \to 180^o$ 
like $\sin^4 (\phi/2)$, would be natural if longitudinal and 
transverse helicity amplitudes have similar strengths but opposite phase 
as may be possible in some simple glueball models. 
\hfill\\
4. The $2^+$ production depends on the dynamics of the meson
as well as the helicity structure of the Pomeron. 
In the non-relativistic $\q\qbar$ model (a particular realization of 
the CVC hypothesis), we predict 
at small $q_{i\perp} \ll M$ a $2^+$ cross section that is 
(i) basically flat in $\cos\phi$, 
(ii) finite for $\kT \to 0$, and 
(iii) dominated by the helicity-two part. 
For the CVC hypothesis a suppression at small $\kT$ is obtained 
only for peculiar relations between the helicity amplitudes. 
Hence again, data show that the CVC Pomeron is not the full story.
In particular, $2^+$ states at $2\,$GeV are seen to have
a different $\phi$ dependence than the established $\q\qbar$ $2^+$ states.

Our analysis can also be applied to $\e\p \to \e\p\, M$ where
$M$ is a vector meson. As $t \to 0$ we find that the longitudinal 
polarization of the meson grows initially as $Q^2/M^2$ relative to
the transverse, with a characteristic $\phi$ dependence.

Readers interested in the results may proceed directly to section~3. 
Their detailed derivation is summarized in section~2.

\section{Derivation of the results}
Consider the central production of a $J^{P+}$ meson $M$ in the 
high-energy scattering of two fermions with momenta $p_1$ and $p_2$,
respectively,
\begin{equation} 
f_1(p_1) + f_2(p_2) \to f_1'(p'_1) + f_2'(p'_2) + M
\ , 
\label{eq:ffM}
\end{equation}
proceeding through the fusion 
of two conserved spin-$1$ vector currents $V_1$ and $V_2$: 
\begin{equation}
  V_1(q_1,\lambda_1) + V_2(q_2,\lambda_2) \to M(J,J_z)
\ .
\label{eq:VVM}
\end{equation}
Here 
$\lambda_i = \pm 1$, $L$
are the current helicities\footnote{%
Our longitudinal-helicity polarization vector $\epsilon_{\mu}(L)$ 
is orthogonal to the momentum vector as are the two 
transverse polarization vectors. Hence in the meson rest frame 
$\epsilon_{\mu}(L)$ has both a $0$ and a $3$ component. Consequently,
our scalar polarization vector is proportional to the momentum vector.}
in the meson rest frame with current one defining the $z$ axis. 
In the case of electron--positron collisions, 
$V_i$ in (\ref{eq:VVM}) is a photon, 
while for central production in proton--proton collisions,
$V_i$ could be a Pomeron, a (colour-less) multi-gluon state, or in some
models, even a single gluon (accompanied by Coulomb gluon(s) to ensure 
colour-conservation). For our purpose here what matters is the assumed
spin-$1$ nature of the production field(s) and their conservation.
We shall comment upon the consequences of current non-conservation 
at the end of the next section. 

In order to investigate the helicity structure of the diffractive agent
it proves useful to examine the dependence of the cross section 
on\footnote{$P$ and $T$ invariance forbid $\sin n\tilde\phi$ contributions.} 
$\cos\, n\tilde\phi$, where $\tilde\phi$ is the azimuthal separation
between the two proton scattering planes of (\ref{eq:ffM}) in the 
current-current c.m.s. An experimental analysis is complicated by two facts.
First, what is measured is not $\tilde\phi$ but the azimuthal angle $\phi$ 
in the proton--proton rest frame. Second, experimental cuts and/or an 
inconvenient choice of kinematical variables might spoil the $\tilde\phi$ 
dependence predicted by theory. 

This is easily understood when one recalls 
that the phase space for (\ref{eq:ffM}), 
$\sim (\d^3p'_1/E'_1)\,  (\d^3p'_2/E'_2)$, 
depends on only four non-trivial variables if the meson is either
stable or has a width much smaller than its mass since one relation is
provided by 
$W \equiv \sqrt{(q_1 + q_2)^2} = M$ (the meson mass). These four variables
are often chosen as four invariants, for example, $Q_i = \sqrt{-q_i^2}$ 
and (suitably defined) fractional current energies $x_i$, or as 
the scattered protons energies and polar angles (or transverse momenta). 
For whatever choice, 
the expression of (the fixed variable) $W$ in terms of these variables 
explicitly involves the angle $\tilde\phi$ (or $\phi$), which introduces 
additional ``spurious'' azimuthal dependences. Moreover, the relation 
between $\tilde\phi$ and the measurable $\phi$ is rather complicated. 

However, as we shall detail below, in the kinematic regime of
experimental interest, we have, to good approximation, $\tilde\phi \approx
\phi$. Also for the WA102 experiment, we estimate the effect of the extra 
kinematic factors to have no significant impact on the effects discussed 
here. 

In the approximation of single-particle (single-trajectory) exchange
(one at each vertex)
the cross section for (\ref{eq:ffM}) factors into the product of three terms,
namely two density matrices and the amplitude for (\ref{eq:VVM}).
Consider the (unnormalized) density matrix for the emission from 
particle $1$. For a conserved vector-current its general form is
\begin{equation}
\rho_1^{\mu\nu} = - 
 \left( g^{\mu\nu} - \frac{ q_1^\mu\, q_1^\nu }{q_1^2} \right)\,
  C_1(q_1^2) - \frac{ (2 p_1 - q_1)^\mu\,  (2 p_1 - q_1)^\nu}{q_1^2}\, 
  D_1(q_1^2)
\ . 
\label{eq:conserved}
\end{equation}
Here $C_1$ and $D_1$ are form factors associated with the non-pointlike 
nature of particle~$1$ (for a lepton, $C_{\e}(q^2) = 1 = D_{\e}(q^2)$, 
while for a proton $C_{\p}(q^2) = G_{\M}^2(q^2)$, 
$D_{\p}(q^2) = (4 m_{\p}^2 G_{\E}^2(q^2) - 
q^2 G_{\M}^2(q^2))/(4 m_{\p}^2 -q^2)$, 
where $G_\E$ and $G_\M$ are the proton electromagnetic form factors). 
A factor $1/(-2 q_1^2)$ in $\rho_1$ is introduced for 
convenience\footnote{With this choice, the matrix elements of the 
first term are simply $\pm C_1$ (or zero), see (\ref{eq:rhodef})
and (\ref{rhoone}).} since current conservation guarantees 
$(2p_1- q_1)^\mu (2p_1 - q_1)^\alpha\, M^{\star\alpha\beta} M^{\mu\nu}
 \propto q_1^2$. 

In the following we shall be working in the current--current 
helicity basis. The density-matrix elements in the helicity basis 
are defined with the help of the polarization vectors 
$\epsilon_1^\mu(\lambda_1)$ of the (space-like) current one as \cite{Budnev}
\begin{equation}
  \rho_1^{\lambda_1,\lambda_1'}
  = (-1)^{\lambda_1 + \lambda_1'}\, 
\epsilon_1^\mu(\lambda_1)\, \rho_1^{\mu\nu}\, 
  \epsilon_1^\nu(\lambda_1')
\ ,
\label{eq:rhohel}
\end{equation}
where $\lambda_1^{(\prime)}$ 
label the helicity of the current one, $\lambda_1^{(\prime)} = \pm 1, L$.
Owing to the hermiticity relations of the density matrix and the
polarization vectors
\begin{eqnarray}
 \rho_1^{\mu\nu\star} & = & \rho_1^{\nu\mu}
\nonumber\\
 \epsilon_1^{\alpha\star}(\pm 1) & = & 
  -  \epsilon_1^{\alpha}(\mp 1) \ , 
 \qquad
 \epsilon_1^{\alpha\star}(L) =  - \epsilon_1^{\alpha}(L) 
\ ,
\label{eq:hermi}
\end{eqnarray}
the helicity-density matrix is determined by four real parameters,
for example, $\rho_1^{++}$,  $\rho_1^{LL}$,  $|\rho_1^{+L}|$, 
and $|\rho_1^{+-}|$. The phases of the latter two matrix elements 
are $\exp(i\tilde\phi_1)$ and $\exp(2 i\tilde\phi_1)$, respectively,
where $\tilde\phi_1$ is the azimuthal angle of $p_1$ in the
current--current c.m.s. (With the analogous definition of $\tilde\phi_2$
we have $\tilde\phi = \tilde\phi_1 + \tilde\phi_2$.)

The expressions of $|\rho_1^{ik}|$ in terms
of invariants and the form factors $C_1$ and $D_1$ can be derived
from the formulas in \cite{Budnev}
\begin{eqnarray}
 \rho_1^{++} & = & C_1 + \frac{1}{2}\, D_1\, \left[
  \frac{ (u_2 - \nu)^2 }{X} - 1 + \frac{4\, m_1^2}{q_1^2} \right]
\nonumber\\
  \rho_1^{LL} & = & - C_1 + D_1\, \frac{ (u_2 - \nu)^2 }{X}
\nonumber\\
  | \rho_1^{+-} | & = & \rho_1^{++} - C_1
\nonumber\\
  | \rho_1^{+L} | & = & 
  \sqrt{ |\rho_1^{+-}|\, \left( \rho_1^{LL} + C_1 \right)}
\ .
\label{eq:rhodef}
\end{eqnarray}
Here we have introduced $u_2 = 2 p_1\cdot q_2$, $\nu = q_1 \cdot q_2
= (W^2 - q_1^2 - q_2^2)/2$, $W^2 = (q_1 + q_2)^2$, 
and $X = \nu^2 - q_1^2\, q_2^2$. 

In this work we are interested in the dominant (and experimentally 
accessible) region of phase space $Q_i \equiv \sqrt{-q_i^2} \ll W$.
Then (and only then \cite{Schuler}) the density matrix 
$\rho_1^{\mu\nu}$ depends on only 
variables of current-one, namely its fractional momentum 
$x_1 = p_2\cdot q_1/ p_2\cdot p_1 = u_1 / (s - 2 m_1^2)$ and 
its virtuality $Q_1$. Moreover, we can use
\begin{eqnarray}
 Q_i & \simeq & q_{i\perp}
\nonumber\\
\tilde\phi & = & \frac{ \vec{q}_{1\perp} \cdot 
    \vec{q}_{2\perp} }{ q_{1\perp}\, q_{2\perp} }
  \simeq \phi = \frac{ \vec{p}^{\ '}_{1\perp} \cdot 
    \vec{p}^{\ '}_{2\perp} }{ p'_{1\perp}\, p'_{2\perp} }
\ ,
\label{Phirel}
\end{eqnarray}
where $\vec{q}_{i\perp}$ ($\vec{p}^{\ '}_{i\perp}$)
is the transverse momentum of current $i$ (scattered proton $i$)
in the current-current (proton--proton) c.m.\ system.
In addition, the dependence of $W = M$ 
on the azimuthal angle $\tilde\phi = \tilde\phi_1 + \tilde\phi_2$ 
disappears, and we simply have $W^2 = x_1 x_2 s$ ($x_2 = u_2/(s - 2 m_2^2))$. 
Since $m_1$ and $m_2$ are much smaller than the
c.m.\ energy $\sqrt{s}$ we obtain
\begin{eqnarray}
 2\, \rho_1^{++} & = & 2\, C_1 + (1 - \delta_1)\, \hat{\rho}_1
\nonumber\\
  \rho_1^{LL} & = & D_1 - C_1 + \hat{\rho}_1
\nonumber\\
 2\, | \rho_1^{+-} | & = & (1 - \delta_1)\, \hat{\rho}_1
\nonumber\\
 \sqrt{2} | \rho_1^{+L} | & = & 
  \sqrt{ (1 - \delta_1)\, \hat{\rho}_1\, \left( D_1 + \hat{\rho}_1 \right)}
\ ,
\label{eq:rhoapprox}
\end{eqnarray}
where we have introduced
\begin{equation}
  \hat{\rho}_1 = \frac{4}{x_1^2}\, \left(1 - x_1\right)\, D_1
\ , \qquad
  \delta_1 = Q^2_{1\min} / Q_1^2
\ .
\label{eq:deltaone}
\end{equation} 

For the production (\ref{eq:ffM}) of mesons at fixed-target experiments 
(and even more so at electron--positron colliders) the meson mass 
is much smaller than the c.m.\ energy. This implies that $x_i \ll 1$
and thus
\begin{equation}
  \frac{2}{1 - \delta_1}\, \rho_1^{++} \simeq 
  \frac{2}{1 - \delta_1}\, |\rho_1^{+-}| \simeq 
  \sqrt{ \frac{2}{1 - \delta_1} }\, |\rho_1^{+L}| \simeq 
      \rho_1^{LL} \simeq \hat{\rho}_1
\ .
\label{eq:rhofinal}
\end{equation}
Relations analogous to (\ref{eq:rhodef})--(\ref{eq:rhofinal}) 
hold also for the density matrix of current two, 
$\rho_2^{\lambda_2,\lambda_2'}$.

Before continuing we have to make sure that (\ref{eq:rhofinal}) is
not spoiled by the behaviour of the form factors, i.e.\ we 
have to make sure that $\hat{\rho}_1 \gg C_1$. This is certainly
true if $C_1 \simeq D_1$ for all $Q_1^2$. To investigate this a bit further
we assume that Pomerons (e.g.\ Pomeron one) couple to fermions 
like the current
\begin{equation}
  J_\mu = \bar{u}(p_1')\, \left\{ F_1(q_1^2)\, \gamma_\mu
  + \frac{\kappa}{2m}\, F_2(q_1^2)\, i \sigma_{\mu\alpha}\, q^\alpha
  \right\}\, u(p_1)
\end{equation}
Then we can actually calculate the density matrix 
(\ref{eq:conserved}) defined by 
\begin{equation}
  \rho_1^{\mu\nu} =   
\frac{-1}{2\, q_1^2}\, \sum_{\mathrm{spins}}\, J_\mu\, J^\star_\nu
\ .
\label{rhoone}
\end{equation}
Noting the minus sign in 
$$ \left( \bar{u}(p_1')\, i \sigma_{\mu\alpha}\, q^\alpha\,  u(p_1)
  \right)^\star = 
   - \bar{u}(p_1)\, i \sigma_{\mu\alpha}\, q^\alpha\,  u(p_1')\ ,$$
we obtain the form (\ref{eq:conserved}) with 
\begin{eqnarray}
  C_1 & = & \left( F_1 + \kappa\, F_2 \right)^2 \equiv G_{\M}^2
\nonumber\\
  D_1 & = & F_1^2 - \frac{q_1^2}{4 m^2}\, (\kappa\, F_2)^2 
   \equiv \frac{4 m^2\, G_{\E}^2 - q_1^2\, G_{\M}^2}{4 m^2 - q_1^2}
\label{Coneresult}
\ .
\end{eqnarray}
Note that a pure $\gamma_\mu$ coupling gives $C_1 = D_1 = F_1^2$.
Hence for two-photon production at $\e^+\e^-$ colliders 
($F_1 = 1$, $F_2 = 0$) our assumptions are well satisfied: 
even at CLEO energies the typical $x_i$ values are small 
enough ($x_i \sim 0.1$) to ensure $1/x_i^2 \gg 1$ and, in turn, 
$\hat{\rho}_1 \gg C_1$. Moreover, the tagging setup of the scattered 
electrons assures that $\delta_i \ll 1$. 

The situation may be different in fixed-target proton--proton collisions.
First, at WA102 energies ($12.8 < \sqrt{s}/\mathrm{GeV} < 28$) 
the experimentally accessible $x_i$ values range between about $10^{-3}$ 
and $0.2$ guaranteeing thus $1/x_i^2 \gg 1$. 
The minimum $x_i$ values result in minimum virtualities 
of $Q^2_{i\min} \approx 10^{-4}\,$GeV$^2$. Hence if we assume that 
measurements are done in a range, say $10^{-3} < Q_i^2/\mathrm{GeV}^2 < 0.5$ 
(statistics limits larger values) then still $\delta_i \ll 1$. 
This holds certainly for the recoil proton since it can only be 
detected for $Q^2$ larger than about $0.05\,$GeV$^2$. The scattered proton
can, however, be measured down to very low $Q^2$. 
For completeness, we shall keep the $(1-\delta_i)$ terms in the following. 

Central production in proton--proton collisions may differ in another 
aspect from the $\e^+\e^-$ case: unlike the photon the Pomeron might have
a dominant $\sigma_{\mu\nu}$-type coupling.
The requirement for (\ref{eq:rhofinal}) to hold, namely 
$\hat{\rho}_1 \gg C_1$, yields for zero $F_1$ the condition
$Q_1^2 \gg Q^2_{1\min}$. Hence as long as very low $Q^2$ values of the 
scattered proton are excluded, (\ref{eq:rhofinal}) continues to hold.
There is one difference, however: if $F_2$ dominates then the typical
$t$ ($t = q^2 = - Q^2$) distribution $\propto \exp(-b t)$ 
(with $b \sim 6/$GeV$^2$) is modified by an extra factor $(-t$).

Let us now continue with the current--current--meson vertex.
If (\ref{eq:VVM}) proceeds through the fusion of
two conserved vector-currents, then conservation of $P$ and $T$ 
as well as total helicity conservation for forward scattering, implies 
that the cross section for 
$f_1 + f_2 \to f_1' + f_2' + X$, for arbitrary final state $X$,
depends on eight independent helicity structure functions, 
$W(\lambda_1,\lambda_2; \lambda_1',\lambda_2' )$
out of which only six can be measured with unpolarized 
initial-state fermions:
\begin{eqnarray}
  \d\sigma & \sim & 
   2\, \rho_1^{++}\, \rho_2^{++}\, 
         \left\{ W(++,++) + W(+-,+-) \right\}
\nonumber\\
&&~ + 2\, \rho_1^{++}\, \rho_2^{LL}\, W(+L,+L) 
\nonumber\\
&&~ + 2\, \rho_1^{LL}\, \rho_2^{++}\, W(L+,L+) 
\nonumber\\
&&~ + \rho_1^{LL}\, \rho_2^{LL}\, W(LL,LL)
\nonumber\\
&&~ + 2\, \left| \rho_1^{+-}\, \rho_2^{+-}\right|\, 
         W(++,--) \, \cos2\, \tilde\phi
\nonumber\\
&&~
    - 4\, \left| \rho_1^{+L}\, \rho_2^{+L}\right|\, 
   \left\{ W(++,LL) + W(L+,-L) \right\} \, \cos\tilde\phi
\ .
\label{eq:difsig}
\end{eqnarray}
Note that $W(\lambda_1,\lambda_2; \lambda_1',\lambda_2' )\neq 0$
only if $\lambda_1 - \lambda_2 = J_z = \lambda_1' - \lambda_2'$. 
Both the structure functions $W$ and the invariant amplitudes $A$ 
defined below in (\ref{eq:Wdef}) are functions of
the invariants $W$, $Q_1^2$, $Q_2^2$ 
only\footnote{If instead one chose to replace one of these variables
by $\phi$ then different $\phi$ dependences could emerge, see, for example,
(5.14) in \cite{Budnev} or (13) in \cite{Johne}.}.

For the present case, (\ref{eq:ffM}), where $X$ is a single particle, 
the number of independent parameters in (\ref{eq:difsig})
can be reduced further. 
First observe that if $A(\lambda_1,\lambda_2)$ denotes
the ($V_1 V_2$ c.m.s.) helicity amplitude for (\ref{eq:VVM}), then we have
\begin{equation}
  W(\lambda_1,\lambda_2; \lambda_1',\lambda_2' )
 = A(\lambda_1,\lambda_2)\,  A^\star(\lambda_1',\lambda_2')\,
  \delta\left( W^2 - M^2 \right)
\ ,
\label{eq:Wdef}
\end{equation}
where $W^2 = (q_1 + q_2)^2$ and $M$ denotes the meson mass.
Second, if $\eta_i$ denotes the naturality\footnote{A boson is said to
have naturality $+1$ if $P=(-1)^J$ and $-1$ if $P=(-1)^{J-1}$.} 
of current $V_i$ and $\eta_M$ that of the meson $M$, then
\begin{equation}
  A(-\lambda_1,-\lambda_2) = \eta\, A(\lambda_1,\lambda_2)
\quad , \quad \eta \equiv \eta_1\, \eta_2\, \eta_M\ ,
\label{eq:etadef}
\end{equation}
and there are five independent helicity amplitudes $A(\lambda_1,\lambda_2)$.
Finally, owing to the $T$-invariance relation
\begin{equation}
  W(\lambda_1,\lambda_2; \lambda_1',\lambda_2' ) =
  W(\lambda_1',\lambda_2'; \lambda_1,\lambda_2 )
\ ,
\label{eq:Tinvar}
\end{equation}
which implies
\begin{equation}
  A(\lambda_1,\lambda_2)\, A^\star (\lambda_1',\lambda_2' ) = 
\mathrm{ csgn}\, A(\lambda_1,\lambda_2)\, 
\mathrm{ csgn}\, A(\lambda_1',\lambda_2')\, 
 \left| A(\lambda_1',\lambda_2') \right| \,
 \left| A(\lambda_1, \lambda_2 ) \right|
\ ,
\label{eq:signs}
\end{equation}
we are left with five real parameters.
Here 
\begin{equation}
  \mathrm{ csgn}\, z = \left\{ \begin{array}{ll} 
        +1 \quad & \mathrm{ Re}\,z > 0 \;\; \mathrm{ or}\; (
            \mathrm{ Re}\,z = 0 \; \mathrm{ and}\; \mathrm{ Im}\, z > 0 ) \\
        -1 \quad & \mathrm{ Re}\,z < 0 \;\; \mathrm{ or}\; (
            \mathrm{ Re}\,z = 0 \; \mathrm{ and}\; \mathrm{ Im}\, z < 0 ) 
\ .
  \end{array}  \right.
\label{eq:csgndef}
\end{equation}

Defining $A_{\lambda_1 \lambda_2} = |A(\lambda_1,\lambda_2)|$ 
and
\begin{eqnarray}
  \xi_1 & = & \mathrm{csgn}\, A(++)\, \mathrm{csgn}\, A(LL)
\nonumber\\
  \xi_2 & = & \mathrm{csgn}\, A(+L)\, \mathrm{csgn}\, A(L+)
\ ,
\label{eq:xidef}
\end{eqnarray}
we find
\begin{eqnarray}
  \d\sigma & \sim & 
   2\, \rho_1^{++}\, \rho_2^{++}\,  A_{+-}^2
\nonumber\\
&&~ + 2\, \rho_1^{++}\, \rho_2^{LL}\, A_{+L}^2   
    + 2\, \rho_1^{LL}\, \rho_2^{++}\, A_{L+}^2
    - 4\, \eta\, 
\left| \rho_1^{+L}\, \rho_2^{+L}\right|\, 
      \xi_2\, A_{+L}\, A_{L+} \, \cos \tilde\phi
\nonumber\\
&&~ + \rho_1^{LL}\, \rho_2^{LL}\, A_{LL}^2 
    - 4\, 
\left| \rho_1^{+L}\, \rho_2^{+L}\right|\, 
      \xi_1\,  A_{++}\, A_{LL} \, \cos \tilde\phi
\nonumber\\
&&~ + 
\left\{ 2\, \rho_1^{++}\, \rho_2^{++} + 2\, \eta\, 
          \left| \rho_1^{+-}\, \rho_2^{+-}\right|\, \cos2\, \tilde\phi
    \right\}\,  A_{++}^2
\ .
\label{eq:ndifsig}
\end{eqnarray}

For the kinematic regime of interest, $x_i \ll 1$ and $Q_1 \ll M$, 
$\tilde \phi \approx \phi$, (\ref{Phirel}), and
(\ref{eq:rhofinal}) allows us to approximate 
\begin{eqnarray}
  \rho_1^{LL}\, \rho_2^{LL} & \approx & 
  \frac{4\, \rho_1^{++}\, \rho_2^{++}}{(1-\delta_1)\,(1-\delta_2)} \approx 
  \frac{2\, \rho_1^{++}\, \rho_2^{LL}}{1-\delta_1} \approx 
  \frac{2\, \rho_1^{LL}\, \rho_2^{++}}{1-\delta_2} 
\nonumber\\ & \approx & 
  \frac{2\, \left| \rho_1^{+L}\, \rho_2^{+L}\right|}%
        {\sqrt{(1-\delta_1)\,(1-\delta_2)}} \approx
  \frac{4\, \left| \rho_1^{+-}\, \rho_2^{+-}\right|}%
        {(1-\delta_1)\,(1-\delta_2)}
\ .
\end{eqnarray}
If we decompose the cross section into components 
(subscript $i$ on $\Sigma_i$) that 
correspond to $|J_z| = 2$, $1$, and $0$, 
then we obtain
\begin{eqnarray}
  \d\sigma & \sim & 
  \Sigma_2 + \Sigma_1 + \Sigma_0
\nonumber\\
  \Sigma_2 & = & 
  \frac{1}{2}\, (1-\delta_1)\,(1-\delta_2)\, A_{+-}^2
\nonumber\\
  \Sigma_1 & = & 
  (1-\delta_1)\, A_{+L}^2   + (1-\delta_2)\, A_{L+}^2
\nonumber\\ &&~ 
    - 2\, \eta\, \xi_2\, \sqrt{(1-\delta_1)\,(1-\delta_2)}\, 
        A_{+L}\, A_{L+} \, \cos \phi
\nonumber\\
  \Sigma_0 & = & 
 A_{LL}^2 
    - 2\, \xi_1\, \sqrt{(1-\delta_1)\,(1-\delta_2)}\, 
       A_{++}\, A_{LL} \, \cos \phi
\nonumber\\ &&~
    + (1-\delta_1)\,(1-\delta_2)\, 
  \left( 1 + \eta\, \cos2\, \phi \right)\, \frac{1}{2} A_{++}^2
\ .
\label{eq:approxsig}
\end{eqnarray}

Introducing
\begin{equation}
  r = \frac{A_{LL}}{A_{++}}
\ ,
\label{eq:rdef}
\end{equation}
and making use of $(1-\eta)\, r = 0$, we can rewrite 
the $J_z = 0$ part in (\ref{eq:approxsig}) as
\begin{eqnarray}
  \Sigma_0 & = &  A_{++}^2 \, \left\{ 
  \delta_{\eta,1}\,
      \left( r - \xi_1 \,  \sqrt{(1-\delta_1)\,(1-\delta_2)}\, 
     \cos \phi \right)^2
\right. \nonumber\\ & &~~~~~ \left.  +~  
  \delta_{\eta,-1}\, (1-\delta_1)\,(1-\delta_2)\, 
    \frac{1}{2} \, \left( 1 - \cos2\, \phi \right)
  \right\}
\ .
\label{eq:zerosig}
\end{eqnarray}
Which of the two terms in (\ref{eq:zerosig}) contributes depends
on the naturality factor $\eta$, see table~\ref{tableone}.
\begin{table}
\begin{center}
\begin{tabular}{|c|c|c|c|c|c|c|c|c|}
\hline
$J^P$ & $0^-$ & $0^+$  & $1^-$ & $1^+$  & $2^-$ & $2^+$  & $3^-$ & $3^+$ 
\\ \hline
$\eta$   & $-$ & $+$  & $+$ & $-$  & $-$ & $+$  & $+$ & $-$ 
\\ \hline
$A_{LL}$ & $0$ & $\delta$  & $D\, \delta$ & $0$  
         & $0$ & $\delta$  & $D\, \delta$ & $0$ 
\\ \hline
$A_{++}$ & $1$ & $1$  & $D$ & $D$  & $1$ & $1$  & $D$ & $D$ 
\\ \hline
$A_{+-}$ & $0$ & $0$  & $0$ & $0$  & $D$ & $1$  & $D$ & $1$ 
\\ \hline
$\kappa$ & $1$ & $0$  & $1$ & $0$  & $1$ & $0$  & $1$ & $0$ 
\\ \hline
\end{tabular}
\caption{Model-independent features of helicity amplitudes up to 
$J^P = 3^+$; 
$0$:~amplitude is identical to zero; 
$D$:~amplitude is proportional to $D = (Q_1^2 - Q_2^2)/M^2$;
$\delta$:~amplitude is proportional to $\delta = Q_1\, Q_2/M^2$ for 
$Q_i \ll M$;
$1$:~amplitude is of order one, in general. 
Also given are the values of $\eta$, 
(\ref{eq:etadef}), and $\kappa$, (\ref{eq:kappadef})
(for the case $\eta_1\, \eta_2 = +1$). 
\label{tableone}}
\end{center}
\end{table}

So far we have not yet made use of Bose
symmetry, which states
\begin{equation}
  A(\lambda_1, \lambda_2)(Q_1,Q_2) = (-1)^J\, 
  A(\lambda_2, \lambda_1)(Q_2,Q_1)
\ ,
\label{Bose}
\end{equation}
where $Q_i = \sqrt{-q_i^2}$ is the virtuality of boson~$i$. It implies 
that (in the CVC hypothesis) 
the amplitudes $A_{++}$ and $A_{LL}$ must be proportional to 
\begin{equation}
 D = \frac{Q_1^2 - Q_2^2}{M^2}
\label{eq:Ddef}
\end{equation}
for odd-integer $J$. When combined with parity, (\ref{eq:etadef}), 
the amplitude $A_{+-} \propto D$ for some $J^P$, 
see table~\ref{tableone}.

Bose symmetry has one more consequence, namely that both
amplitudes $A_{L+}$ and $A_{+L}$ in (\ref{eq:approxsig}) can be replaced
by only one of them, say $A_{+L}$. Moreover, the sign in (\ref{eq:signs})
is then fixed in a model-independent way. We 
can rewrite the $|J_z|=1$ part of the cross section as
\begin{eqnarray}
 \Sigma_1 & = & 
   (1-\delta_1)\,  A_{+L}^2(Q_1,Q_2) + (1-\delta_2)\, A_{+L}^2(Q_2,Q_1)  
\nonumber \\ &&
     - 2\, (1 - 2\, \kappa)\, \sqrt{(1-\delta_1)\,(1-\delta_2)}\, 
     A_{+L}(Q_1,Q_2)\, A_{+L}(Q_2,Q_1)\, \cos \phi
\ ,
\label{eq:onesig}
\end{eqnarray}
where we have introduced the variable
\begin{equation}
  \kappa = \frac{1 - \eta\, (-1)^J }{2}
\ ,
\label{eq:kappadef}
\end{equation}
whose values, one or zero, are given in table~\ref{tableone} for 
states up to $J^P = 3^+$. 

We can exploit one more constraint, namely current conservation, which
requires
\begin{eqnarray}
  A_{\pm 1,L} \propto & Q_2/M               & \mathrm{for}\quad Q_2 \ll M
\nonumber\\
  A_{L,\pm 1} \propto & Q_1/M               & \mathrm{for}\quad Q_1 \ll M
\nonumber\\
  A_{L,L}     \propto & Q_1\, Q_2/M^2 \quad & \mathrm{for}\quad Q_i \ll M
\label{gauge}
\ .
\end{eqnarray}
Then (\ref{Phirel}) implies that
\begin{eqnarray}
  A_{+L} & \simeq & a_{+L}\, \frac{ q_{2\perp} }{M}
\nonumber\\
  A_{LL} & \simeq & a_{LL}\, \delta\ , 
   \qquad \delta \equiv \frac{Q_1\, Q_2}{M^2} 
       \simeq \frac{q_{1\perp}\, q_{2\perp}}{M^2} 
\nonumber\\
D & \simeq & \frac{ q_{1\perp}^2 - q_{2\perp}^2 }{M^2}
\ ,
\label{eq:Dapprox}
\end{eqnarray} 
where $a_{ij}$ are coefficients of order one.
Hence $\Sigma_1$ in (\ref{eq:onesig}) behaves as
\begin{equation}
  \Sigma_1 = \left\{ \begin{array}{lll}
       a_{+L}^2\, \pT^2 / M^2\, , & \mathrm{for}\; \kappa = 1\  , \\
       a_{+L}^2\, \kT^2 / M^2\, , & \mathrm{for}\; \kappa = 0
\  ,
                     \end{array}
  \right.
\label{eq:oneapprox}
\end{equation}
where
\begin{eqnarray}
\pT^2  & = & 
 \left( \sqrt{1-\delta_2}\, \vec{q}_{1\perp} + 
        \sqrt{1-\delta_1}\, \vec{q}_{2\perp} \right)^2 
\nonumber\\
\kT^2 & = & 
 \left( \sqrt{1-\delta_2}\, \vec{q}_{1\perp} - 
        \sqrt{1-\delta_1}\, \vec{q}_{2\perp} \right)^2 
\ .
\label{eq:kTdef}
\end{eqnarray}
Note that 
$\kT \to 0$ implies $\phi \to 0$ {\em and} $q_{2\perp} \to q_{1\perp}$. 
However, the opposite is not true: 
$\phi \to 0$ does not in general imply $\kT \to 0$. 

\section{Results}
The above analysis enables some immediate conclusions to be drawn
according to the $J^{PC}$ of the meson.\hfill\\
\noindent
{\bf (i)} $\bJ^\bP \beq \bzero^\bminus$\hfill\\
Only $J_z=0$ contributes and, with $\eta=-1$ in (\ref{eq:zerosig})
\begin{equation}
\frac{\d \sigma}{\d \phi} \propto
  A_{++}^2\, (1 - \delta_1)\, (1 - \delta_2)\, \frac{1}{2}\, \left(
   1 - \cos 2\, \phi \right)
  = 
  A_{++}^2\, (1 - \delta_1)\, (1 - \delta_2)\,  \sin^2 \phi
\ .
\label{eq:spinzerominus}
\end{equation}
This follows independent of the dynamical internal structure of the
$0^{-+}$ meson, and is simply a consequence of parity.
Since $\phi \to 0$ as $\kT \to 0$ 
we recover the result of \cite{Close,Castoldi,Frere} 
who noted that the production of $0^{-+}$
by (conserved) vector currents would vanish as $\kT \to 0$. 
Our result above provides a clear test for the vector nature of the 
production Pomeron (component) by the explicit prediction for
the $\phi$ dependence, independent of the $t$-dependence.

Preliminary indications are that the production of $\eta$ and $\eta'$ in
$\p\p \to \p\p\eta$ ($\eta'$) is compatible with such a $\phi$ dependence 
\cite{AKprivate}. 

\noindent
{\bf (iii)} $\bJ^\bP \beq \bone^\bplus$\hfill\\
Since $|J_z| \leq 1$ the azimuthal distribution is given by the sum
of (\ref{eq:zerosig}) (with $\eta = -1$) and (\ref{eq:onesig}) 
(with $\kappa = 0$). 
Since Bose symmetry yields $A_{++} = a_{++}\, D$ we find with
the help of table~\ref{tableone} and (\ref{eq:Dapprox})
in the region of small $Q_i$ 
\begin{equation}
  \frac{\d\sigma}{\d\phi} \sim a_{+L}^2\, \frac{\kT^2}{M^2}
   + a_{++}^2\, (1-\delta_1)\, (1-\delta_2)\, 
   \sin^2 \phi\, 
    \frac{ \left( q_{1\perp}^{2} - q_{2\perp}^{2} \right)^2}{M^4}
\ .
\label{eq:spinone}
\end{equation}
From this we can draw conclusions, which are independent of the 
internal structure of the $1^{++}$ meson and thus hold for
both $\e^+\e^-$ collisions and diffractive proton--proton collisions 
mediated by a vector Pomeron. First, the cross section will tend to 
zero as $\kT \to 0$. 
And second, $1^+$ mesons are produced dominantly in
the helicity-one state. 
Both of these phenomena are seen in the central production of 
$1^{++}$ mesons in $\p\p$ collisions 
which further supports the importance of the
vector component of the effective Pomeron.

The tendency for large $\kT$ to correlate with large $\phi$ may cause 
the apparent $\d\sigma / \d\phi$ to rise as $\phi \to 180^o$. 
The $\phi$ distributions should be binned in $\kT$ to extract the
full implications of (\ref{eq:spinone}).

\noindent
{\bf (ii)} $\bJ^\bP \beq \bzero^\bplus$\hfill\\
In this case the $\phi$ dependence depends on 
the internal structure of the meson and dynamics, 
specifically via the magnitude of $A_{LL} / A_{++} \equiv r$
\begin{equation}
\frac{\d \sigma}{\d \phi} = 
  A_{++}^2\, 
  \left( \sqrt{(1 - \delta_1)\, (1 - \delta_2)}\, 
       \xi_1\, \cos \phi - r \right)^2
\ .
\label{eq:spinzeroplus}
\end{equation}
At small $Q_i$, $Q_i \ll M$ in $\e^+\e^-$ collisions, we have 
$r \simeq c\, \delta 
\simeq c\, q_{1\perp}\, q_{2\perp}/M^2$ 
with $c = a_{LL}/A_{++} = O(1)$, in general.

For the particular case of two photons coupling to a non-relativistic 
quark-antiquark one has \cite{Gulik,Galuga} $\xi_1 = +1$ and $c=4/3$ since
\begin{equation}
 r = \frac{ Q_1\, Q_2 \, M^2 }{\nu^2 + \nu\, M^2 - Q_1^2\, Q_2^2}
 \approx \frac{4}{3}\, \frac{ q_{1\perp}\, q_{2\perp} }{M^2}
\end{equation}
at $q_{i\perp} \ll M$. 
Hence for tagged two-photon events in $\e^+\e^-$ collisions we predict 
a cross section that survives the $\kT \to 0$ limit and 
the $\phi$ distribution (\ref{eq:spinzeroplus}), which 
for $q_{i\perp} \ll M$ is a pure $\cos^2\phi$ distribution. 

This will also hold true for $\q\qbar$ {\em and} glueball production 
in $\p\p$ collisions {\em if} the Pomeron is a {\em conserved} 
vector current. So far we have taken the simplest assumption needed
for a nontrivial $\phi$ distribution, namely CVC. 
This is immediately relevant to $\e^+\e^-$ but encouragingly
shows consistency with $\p\p$. The $0^-$ is a direct test 
with its $\sin^2\phi$ distribution which is verified for $\eta$, 
$\eta'$ in WA102. For $1^+$ the $\kT \to 0$ vanishing and the
helicity-$1$ dominance are verified. The $0^+$, $2^+$ data
clearly go beyond this. 

The non-trivial $\phi$ dependence required 
$J_{\mathrm{Pomeron}} > 0$ to be present 
but leaves open the question of whether there is a spin-$0$ 
component in addition to the CVC and/or whether there is a
non-conserved vector current. Note that the $0^-$ production 
is not sensitive to any $0^+$ component in the Pomeron. 
The simplest manifestation of a scalar component or a non-conserved 
vector piece, is to allow $R$ to be larger than its CVC suppression 
$O(\sqrt{t_1 t_2}/M^2)$. The $0^+$, $2^+$ data are consistent with this. 
The $\f_0(1500)$ production, in particular, is well described 
if $R$ is negative with $|R|\sim O(1)$, in which case 
its $\phi$ distribution is $\sim \sin^4 (\phi/2)$. 
This sign and magnitude are natural for the production of a 
gluonic system if the dynamics for $M_{LL}/M_{++}$ is driven
by the Clebsch--Gordon coefficients
$\langle 10,10 | 00 \rangle / \langle 11,1-1 | 00 \rangle = -1$.
We leave the discussion of the phenomenology and specific models
to a later publication.

\noindent
{\bf (iii)} $\bJ^\bP \beq \btwo^\bplus$\hfill\\
The azimuthal distribution is given by the sum of 
$\Sigma_2$, (\ref{eq:approxsig}), 
$\Sigma_1$, (\ref{eq:onesig}) with $\kappa = 0$, and
$\Sigma_0$, (\ref{eq:zerosig}) with $\eta = +1$. 
Using the small-$Q_i$ approximation for $\Sigma_1$ we have 
\begin{equation}
  \d\sigma \sim \frac{1}{2}\, A_{+-}^2 \, (1-\delta_1)\, (1-\delta_2)\, 
    + a_{+L}^2\, \frac{\kT^2}{M^2}
   + \left( r - \xi_1\,\sqrt{ (1-\delta_1)\, (1-\delta_2)}\, 
   \cos\, \phi \right)^2\, A_{++}^2
\ .
\label{eq:spintwoplus}
\end{equation}
As we can see the $|J_z| = 1$ part is suppressed as is $A_{LL}$
(recall $r \sim \delta$ at small $Q_i$). However, 
in general, the other two amplitudes are of order one, i.e.\ 
$A_{+-} \sim A_{++} \sim 1$. 

In the non-relativistic quark model, 
$A_{++} \simeq (Q_1^2 + Q_2^2)/M^2$ at small $Q_i$ 
\cite{Li,Gulik} and is thus
very much suppressed relative to $A_{+-}$, which is $O(1)$. 
Hence in $\e^+\e^-$ collisions at small $q_{i\perp} \ll M$ 
we predict a $2^+$ cross section that is 
(i) basically flat in $\cos\phi$, 
(ii) finite for $\kT \to 0$, and 
(iii) dominated by the helicity-two part. 
We necessarily obtain the same behaviour, namely
flat $\phi$ distribution and $\kT \to 0$ survival, 
in diffractive $\p\p$ collisions mediated by a conserved vector Pomeron, 
provided the helicity-two component is the dominant one. 

If the Pomeron--$\q\qbar$ coupling were dominantly ``magnetic'' 
(flipping the spins of the produced $\q\qbar$ pair but leaving 
them in an $L_z = 0$ state) the helicity-two amplitude $A_{+-}$ would
be suppressed. In this case the helicity-one amplitude would also
be suppressed as $\kT \to 0$ and the helicity-zero amplitude 
would dominate with a characteristic $\phi$ dependence 
(unless $A_{++} = 0$). Moreover, the $2^+$ cross section
continues to survive the $\kT \to 0$ limit since $r$ is small for CVC.

Again we conclude that, as for $0^+$ production, a non-conserved 
vector piece (or a large scalar component) is needed to accommodate
for the observed small-$\kT$ suppression of $\f_2(1270)$ and
$\f'_2(1520)$. In the scenario discussed above this follows if 
$\xi_1\, r \sim O(1)$. We point out that
these predictions assume Pomeron--Pomeron or gluon--gluon fusion 
and hence do not apply to $\f_2$ production if the latter
has a substantial contribution from $\f_2$ exchange (i.e.\ from 
$\f_2 +\, $Pomeron $\rightarrow$ $\f_2$. A detailed comparison 
of $\f_2$ $\s\sbar(1525)$ (for which this contribution is suppressed)
and $\f_2$ $\u\ubar(1270)$ (where Pomeron--$\f_2$ is possible) could 
help settle this. 

\noindent
{\bf (iii)} $\bJ^\bP \beq \btwo^\bminus$\hfill\\
Here we find with the help of table~\ref{tableone} and (\ref{eq:Dapprox})
\begin{equation}
  \d\sigma \sim (1-\delta_1)\, (1-\delta_2)\, 
  \left\{ \frac{1}{2}\, a_{+-}^2 \, 
    \frac{ \left( q_{1\perp}^{2} - q_{2\perp}^{2} \right)^2}{M^4}
   + \sin^2\phi \, A_{++}^2 
  \right\}
   + a_{+L}^2\, \frac{\pT^2}{M^2}
\ .
\label{eq:spintwominus}
\end{equation}
The helicity-two component vanishes as $\kT \to 0$, as does 
the helicity-zero also. However, the helicity-one component 
($\propto \pT^2$) 
stays non-zero, in general. In the quark model coupling to two photons, 
both $a_{+-}$ and $a_{+L}$ are zero \cite{Li,Gulik}, and so in 
this model the 
cross section will have the same features as that of a $0^{-+}$ meson, 
namely a $\phi$ distribution 
$\propto \sin^2\phi$, 
a cross section that vanishes for $\kT \to 0$, and helicity-zero dominance. 

For central production in hadronic reactions mediated by a vector Pomeron
we have to distinguish two cases, namely $a_{+L} \neq 0$ or $= 0$. 
In both cases the helicity-two component is suppressed. 
In the first case we have 
a cross section that survives at small $\kT$. Moreover, at small $\kT$ 
we expect helicity-one dominance and a flat $\phi$ distribution. 
In the second case, i.e.\ for a suppressed helicity-one amplitude, 
we predict (i) a vanishing $2^-$ cross section for $\kT \to 0$ (recall, 
both $q_{1\perp}-q_{2\perp}$ and 
$\sin^2\phi$ 
vanish for $\kT \to 0$), 
and, provided $A_{++} \neq 0$, helicity-zero dominance as well as a
$\sin^2\phi$ 
distribution (since 
$(q_{1\perp}^2-q_{2\perp}^2)^2/M^4$ is smaller at low $\kT$ than
$\sin^2\phi$).

\noindent
{\bf (iv)} $\bJ^\bP \beq \bthree^\bplus$ {\bf and}
$\bone^\bminus$, $\bthree^\bminus$\hfill\\
With the help of table~\ref{tableone} and (\ref{eq:Dapprox}) 
it is straightforward to find the $\kT$ and $\phi$ distributions for the
$3^+$ states and possible (non-$\q\qbar$) $1^{-+}$ and $3^{-+}$ states.

\begin{eqnarray}
  \d\sigma[3^+] & \sim & 
  \frac{1}{2}\, A_{+-}^2 \, (1-\delta_1)\, (1-\delta_2)\, 
    + a_{+L}^2\, \frac{\kT^2}{M^2}
\nonumber\\ & &~
   + D^2\, a_{++}^2 \, (1-\delta_1)\, (1-\delta_2)\, 
   \sin^2\phi
\nonumber\\
  \d\sigma[3^-] & \sim & 
  \frac{1}{2}\, D^2\, a_{+-}^2 \, (1-\delta_1)\, (1-\delta_2)\, 
    + a_{+L}^2\, \frac{\pT^2}{M^2}
\nonumber\\ & &~
   + D^2\, a_{++}^2 \, 
       \left( \delta\, \frac{a_{LL}}{a_{++}}
         - \xi_1\,\sqrt{ (1-\delta_1)\, (1-\delta_2)}\, 
              \cos\, \phi \right)^2
\nonumber\\
  \d\sigma[1^-] & \sim & 
    a_{+L}^2\, \frac{\pT^2}{M^2}
    +   D^2\, a_{++}^2 \, 
       \left( \delta\, \frac{a_{LL}}{a_{++}}
         - \xi_1\,\sqrt{ (1-\delta_1)\, (1-\delta_2)}\, 
              \cos\, \phi \right)^2
\ .
\label{eq:spinminus}
\end{eqnarray}
Here we have used that $A_{LL} = D \delta a_{LL}$ for $1^-$ and $3^-$ mesons. 

As a specific example\footnote{A word of caution is appropriate: So far
we have not used conservation of charge conjugation; independent 
of $C$[Pomeron], the meson is $C=+1$. In order for the $\e\p$ application
to hold we assume in the following the Pomeron to be $C=+1$ although it couples
like a $C=-1$ photon.}
we illustrate the $1^{--}$ which can be most
immediately relevant in $\e\p \to \e \p\, V$. 
Note that Bose symmetry is now not valid and so both the form of
the first term in (\ref{eq:spinminus}) is changed and the factor
$D^2$ is absent in the second term. We find
\begin{eqnarray}
  \d\sigma & \sim & 
 \left( \sqrt{1-\delta_2}\, \frac{a_{L+}}{M}\, \vec{q}_{1\perp} 
    - \eta\, \xi_2\, 
        \sqrt{1-\delta_1}\, \frac{a_{+L}}{M}\, \vec{q}_{2\perp} \right)^2 
\nonumber\\ & &~
   + A_{++}^2\, 
  \left( r - \xi_1\, \sqrt{ (1-\delta_1)\, (1-\delta_2) }\, 
           \cos\phi \right)^2 
\ .
\label{eq:spinoneminus}
\end{eqnarray}
In the particular case of forward electroproduction, where $t_2 \to 0$
but $q_{1 \perp}^2 = Q^2$ is small, we have approximately
\begin{equation}
 \d\sigma \sim \frac{Q^2}{M^2}\,  a_{L+}^2
   + 
     A_{++}^2\,
     \left( r - \cos\phi \right)^2
\ .
\end{equation}
If for some reason we still have $A_{++} = D\, a_{++}$ or
if $A_{++}(Q_1,Q_2=0) \sim Q_1^2$ then
\begin{equation}
 \d\sigma \sim  a_{L+}^2
   + \frac{Q^2}{M^2}\, 
     \left( r - \cos\phi \right)^2
\ .
\end{equation}
Thus we would expect dominance of transversely-polarized vector-mesons 
and a longitudinally-polarized component of a characteristic 
$\phi$ dependence.

In this section we have given explicit formulae for the CVC case only.
While this applies to $\e^+\e^- \to \e^+\e^- M$, we have noted that
some data involving the Pomeron in proton--proton collisions go beyond
this hypothesis. We will discuss elsewhere the detailed phenomenology 
for both $\p\p$ and $\e\p$-induced reactions. 

\noindent
{\it Acknowledgement}\hfill\\
It is a pleasure to thank A.\ Donnachie, J.\ Ellis, 
W.\ Hollik, J.-M.\ Fr\`{e}re, and A.\ Kirk for useful discussions.

\end{document}